\documentclass[usenatbib]{mn2e}
\usepackage{epsfig}
\usepackage{natbib}
\usepackage{bibentry}
\usepackage{amsmath}
\usepackage{amssymb}

\graphicspath{{Figs/}}

\newcommand{\aap}{A\&A}

\newcommand{\apjl}{ApJ}
\newcommand{\apjs}{ApJS}

\newcommand{\etal}{et~al.}
\newcommand{\ionhy}{H{\sc ii}}

\newcommand{\kms}{$\mbox{km~s}^{\scriptsize{-1}}$}
\newcommand{\msol}{\mbox{$M$\hbox{$_\odot $}}}
\newcommand{\lsol}{\mbox{$L$\hbox{$_\odot $}}}

\begin{document}

\title[Parallaxes of 6.7--GHz Methanol Masers towards G\,305.2]{Parallaxes of 6.7--GHz Methanol Masers towards the G\,305.2 High--Mass Star Formation Region}
\author[Krishnan \etal]{V.~Krishnan,$^{1,2,3}$\thanks{Email: vasaantk@arcetri.astro.it}
                     S.~P.~Ellingsen,$^{2}$
                     M.~J.~Reid,$^{4}$
                     H.~E.~Bignall,$^{3,5}$
                        J.~McCallum,$^{2}$ \newauthor
                     C.~J.~Phillips,$^{3}$
                        C.~Reynolds,$^{3,5}$
                        J.~Stevens,$^{3}$ \\
  $^1$ INAF--Osservatorio Astrofisico di Arcetri, Largo E. Fermi 5, 50125 Firenze, Italy\\
  $^2$ School of Mathematics and Physics, University of Tasmania, Private Bag 37, Hobart, Tasmania 7001, Australia\\
  $^3$ CSIRO Astronomy and Space Science, Australia Telescope National Facility, CSIRO, PO Box 76, Epping, NSW 1710, Australia\\
  $^4$ Harvard--Smithsonian Center for Astrophysics, Cambridge, Massachusetts 02138, USA\\
  $^5$ International Centre for Radio Astronomy Research, Curtin University, Building 610, 1 Turner Avenue, Bentley WA 6102, Australia}
 \maketitle

\begin{abstract}
We have made measurements to determine the parallax and proper motion of the three~6.7--GHz methanol masers G\,305.200$+$0.019, G\,305.202$+$0.208 and G\,305.208$+$0.206. The combined parallax is found to be~0.25$\pm $0.05~mas, corresponding to a distance of~4.1$^{+1.2}_{-0.7}$~kpc. This places the G\,305.2 star formation region in the Carina--Sagittarius spiral arm. The inclusion of G\,305.2 increases the Galactic azimuth range of the sources in this arm by~40$^\circ $ from \citeauthor{Sato+14}, allowing us to determine the pitch angle of this spiral with greater confidence to be~$\psi = 19.0 \pm 2.6^\circ $. The first VLBI spot maps of the~6.7--GHz methanol masers towards these sources show that they have simple linear and ring--like structures, consistent with emission expected from class~II methanol masers in general.
\end{abstract}

\begin{keywords}
masers -- stars: formation -- Galaxy: structure
\end{keywords}


\section{Introduction}
\label{sec:g305-introduction}

The northern hemisphere VLBI telescopes, including the Very Long Baseline Array~(VLBA), VLBI Exploration of Radio Astrometry~(VERA) and European VLBI Network~(EVN) arrays are currently involved in programs to determine the parallaxes to high--mass star formation regions~(HMSFRs) in the Milky Way, by measuring the relative separation between maser emission associated with these regions and distant background quasars. The sub--milliarcsecond~(mas) accurate astrometric measurements between the maser spots and quasars allow the small trigonometric parallax signatures to be detected, and the corresponding distances to these Galactic HMSF masers can be determined to an accuracy of~10\% at~10~kpc \citep{Reid+14b}.

Using over~100~parallax measurements to HMSFRs in the Milky Way, \citet{Reid+14a} have determined the latest Galactic rotation and dynamical parameters, finding the circular rotation speed of the Sun to be~$\Theta _0 = 240 \pm 8 $~\kms \/ and distance to the Galactic centre to be~$R_0 = 8.34 \pm 0.16$~kpc. These results have been obtained from measurements primarily from the first and second quadrants of the Galaxy, and in \citet{Reid+16} the authors demonstrate that the complete spiral structure of the Milky Way cannot be clearly distinguished without parallax distances from southern hemisphere HMSFRs. In~2008 we initiated a project to observe~6.7--GHz class~II methanol masers in the southern hemisphere using the Australian Long Baseline Array~(LBA) for parallax determination.

The methanol molecule has a rich radio and millimetre wavelength spectrum \citep[e.g.][]{Muller+04}, with more than~30~observed transitions detected in interstellar space. Many of these transitions are observed to exhibit maser emission, which are empirically grouped into two classifications \citep{Menten91a}. Class~I methanol transitions are associated with distant parts of the outflows \citep{Cyganowski+09,Voronkov+06} or other shocks \citep{Voronkov+10a} and class~II masers are associated close to the young star at distances of around~10~--~10$^3$~AU \citep[e.g.][]{Sanna+10b,Ellingsen+06}. The~6.7~and~12.2~GHz~class~II methanol masers are two of the strongest and best studied transitions of astrophysical maser emission with the~12.2~GHz masers forming a complete subset of the~6.7~GHz emission \citep{Breen+12a,Caswell+95}. The~6.7~GHz maser is a particularly important transition, as it is exclusively observed towards HMSFRs \citep{Breen+13}. Class~II methanol masers are pumped by radiative excitation of the methanol molecule \citep{Sutton+01} and are known to be strong radio sources with individual features exhibiting point--like structure -- even at VLBI resolution -- making them excellent candidates for astrometry.

The LBA observations continue to be the only southern hemisphere astrometric measurements of Galactic HMSFR masers for parallax determination. The first parallax distance to a southern methanol maser source has been presented in \citet{Krishnan+15}, and here we provide trigonometrical parallax distances to the~6.7--GHz methanol maser sources in the G\,305.2 star formation region.

\section{The G\,305.2 Complex}
\label{sec:g-305-complex}

\begin{figure}
 \centering
 \includegraphics[width=\linewidth]{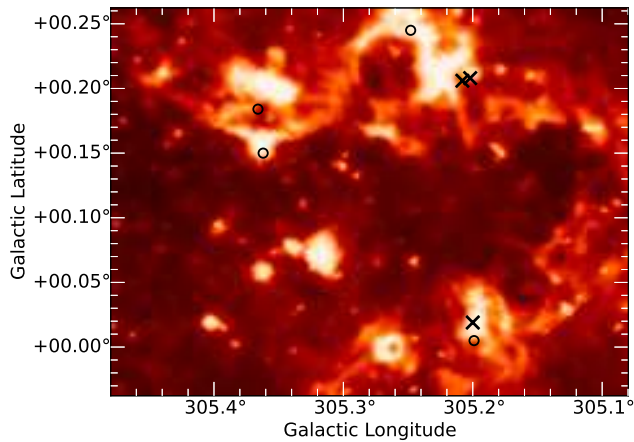}
\caption{6.7--GHz methanol maser emission associated with the G\,305.2 HMSFR superimposed on an infra--red image from \emph{Spitzer} with data at~3.6 and~4.5~$\mu $m. The crosses indicate the positions of G\,305.200$+$0.019, G\,305.202$+$0.208 and G\,305.208$+$0.206 which were observed between~2013 and 2015~March. Positions of the other~6.7--GHz methanol masers in the vicinity from \citet{Green+12a} are shown as circles.}
\label{fig:spitzer}
\end{figure}

The G\,305.2 region is a vast HMSFR in the southern Galactic Plane, with extensive studies undertaken by \citet{Hindson+13,Hindson+12,Hindson+10,Davis+12,Faimali+12,Walsh+07,Walsh+02,Walsh+06,Clark+04} amongst others. The sources which exhibit~6.7--GHz methanol maser emission, include G\,305.200$+$0.019 in the south west and G\,305.202$+$0.208 and G\,305.208$+$0.206 in the north west as described in Table~\ref{tab:CO-clumps} and Figure~\ref{fig:spitzer}. \citet{Phillips+98} published the~6.7--GHz methanol maser spectrum and component map for G\,305.202$+$0.208 (as G\,305.202$+$0.207) and report a curved distribution of maser emission with a peak of~92~Jy at~$-$44.0~\kms . ATCA observations by \citet{Norris+93} show G\,305.208$+$0.206 to have a linear distribution of~6.7--GHz methanol maser spots, with a monotonic velocity gradient and a peak of~$>$300~Jy at~$-$38.3~\kms . The G\,305.208$+$0.206 site contains both~OH and methanol masers \citep{Caswell+95a}, and is located only~22\arcsec \/ to the east of G\,305.202$+$0.208. The~6.7--GHz methanol maser emission towards G\,305.200$+$0.019 is weaker than the emission associated with either G\,305.202$+$0.208 or G\,305.208$+$0.206, with a peak flux density of~46~Jy at~$-$33.1~\kms \/ \citep[][methanol multibeam~(MMB) catalogue]{Green+12a}.

Since the~1980s there have been numerous attempts to determine the distance to the G\,305.2 complex using various techniques, resulting in a range of values between~2.8~--~6.2~kpc \citep[e.g.][]{Russeil+98,Phillips+98,Walsh+97,Caswell+87a,Danks+83}.
The uncertainty in the derived properties of the young high--mass stars in G\,305.2 is predominantly due to the range of the adopted distances by various authors. However the discrepancy is also influenced by the observations, for example, by the arcminute sized beams of the Infrared Astronomical Satellite~(\emph{IRAS}) and Parkes observations in the analysis by \citet{Walsh+97}, compared to the~$\sim $10\arcsec \/ resolution of the \citet{Walsh+07} observations. \citet{Walsh+97} use a distance of~2.8~kpc to G\,305.200$+$0.019 (which they identify as G\,305.202$+$0.019) to derive a luminosity of~$13.2\times 10^4$~\lsol \/ for the associated~\emph{IRAS} source, corresponding to an O\,6.5 spectral type. In contrast, \citet{Hindson+12} use a distance of~3.8~kpc to classify G\,305.200$+$0.019 as a~B\,1~source using the intensity of the associated UC\ionhy \/ region. \citet{Walsh+07} confirm the hot core nature of G\,305.208$+$0.206 through their molecular observations of CH$_3$OH, CH$_3$CN, NH$_3$, OCS and H$_2$O, and estimate the age of the core to be between~$2.0\times 10^4$ to~$1.5\times 10^5$~yr. Unlike G\,305.202$+$0.208 \citep[referred to as G\,305B by][]{Walsh+01} which is associated with a very bright and reddened~IR~source \citep{DeBuizer+03,Walsh+99}, there is little evidence to suggest coincident~IR~emission associated with G\,305.208$+$0.206~\citep[referred to as G\,305A by][]{Walsh+01}. Based on the absence of detectable \ionhy \/ regions, \citet{Walsh+06} propose that G\,305.202$+$0.208 and G\,305.208$+$0.206 are at early stages of stellar evolution, with G\,305.208$+$0.206 likely to be the younger source, as it shows strong mm--continuum emission \citep{Csengeri+14,Hill+05} and no detectable~IR~emission. Assuming a distance of~6.2~kpc, \citet{Phillips+98} propose upper--limit spectral types of~$<$B\,1 and~$<$B\,0.5 for G\,305.208$+$0.206 and G\,305.202$+$0.208, derived from an absence of associated UC\,\ionhy \/ regions (with a detection limit of~$\sim $0.5~mJy~beam$^{-1}$ at~8.6--GHz).

Table~\ref{tab:CO-clumps} describes the properties of the most massive~$^{13}$CO~molecular clump from \citet{Hindson+13} which is associated with G\,305.200$+$0.019, G\,305.202$+$0.208 and G\,305.208$+$0.206. Using Galactic parameter values of~$\Theta _0 = 240$~\kms \/ (including an uncertainty of~8~\kms ) and~$R_0 = 8.34$~kpc \citep{Reid+14a}, we find the~(near) kinematic distance to the G\,305.2 region to be~4.3$^{+2.2}_{-1.4}$~kpc. A known caveat of the kinematic distance technique is that it is susceptible to large error if the sources exhibit anomalous peculiar motions \citep[e.g.][]{Xu+09}. Table~\ref{tab:CO-clumps} shows that there is a difference of~$\sim $10~\kms \/ between the~MMB catalogue peak velocity of the~6.7--GHz methanol maser emission and the associated~$^{13}$CO~clump G\,305.21$+$0.21. Therefore using the kinematic distance technique can produce results with considerable error for the maser sources in this region.

As a prominent southern star formation region, improved accuracy in the distance determination to the G\,305.2 complex is clearly significant, and here we report the trigonometric parallax distance to this complex.

\begin{table}
  \caption{A comparison of the~$^{13}$CO~molecular clump from \citet{Hindson+13} and the~6.7--GHz methanol masers in G\,305.200$+$0.019, G\,305.202$+$0.208 and G\,305.208$+$0.206 \citep[source $v_{\mbox{\scriptsize lsr}}$ from; ][]{Green+12a}. The separation column describes the angular distance between the~6.7--GHz methanol maser from the clump. G\,305.21$+$0.21 has a spread of~$\Delta V = 4.46$~\kms \/ in the~$v_{\mbox{\scriptsize lsr}}$ and a mass of~6300~\msol .}
  \centering
  \begin{tabular}{rlcc}
\hline
 & \multicolumn{1}{c}{\bf Source} & \bf $v_{\mbox{\scriptsize lsr}}$ & \bf Separation \\
 & \multicolumn{1}{c}{\bf name}   & \bf (\kms )                 & \\
\hline
\hline
$^{13}$CO~clump: &                    &              &             \\
            &  G\,305.21$+$0.21      & $-$43.3      &  --         \\
Masers:     &                        &              &             \\
            &  G\,305.200$+$0.019    & $-$33.1      & ~0.2\arcmin \\
            &  G\,305.202$+$0.208    & $-$44.0      & 29.5\arcsec \\
            &  G\,305.208$+$0.206    & $-$38.3      & 15.2\arcsec \\
\hline
  \end{tabular}
  \label{tab:CO-clumps}
\end{table}

\section{Observations}
\label{sec:observations-II}

\begin{table*}
\caption{Phase--referenced 6.7--GHz methanol maser observations between~2013 and 2015~March including the start time and the duration of the observations. The participating stations are the Australia Telescope Compact Array~(AT), Ceduna~(CD), Hartebeesthoek~(HH), Hobart~(HO), Mopra~(MP), Parkes~(PA) and Warkworth~(WA).}
\begin{tabular}{rccccl}
\hline
& & & \multicolumn{1}{c}{\bf UT} & \multicolumn{1}{c}{\bf Duration} & \multicolumn{1}{c}{\bf Participating} \\
\bf Year& \bf Date & \bf D.O.Y & \bf Start & \bf (Hours) & \multicolumn{1}{c}{\bf stations} \\
\hline
\hline
 2013 & 18 Mar & 077 & 04:00 & 24   & AT, CD, HH, HO, PA         \\
 2013 & 17 Jun & 168 & 02:30 & 19.5 & AT, CD, HH, HO, MP, PA     \\
 2013 & 14 Aug & 226 & 18:00 & 24   & AT, CD, HH, HO, MP, PA     \\
 2013 & 19 Nov & 323 & 12:00 & 23   & AT, CD, HO, MP, PA         \\
 2015 & 27 Mar & 086 & 02:00 & 24   & AT, CD, HH, HO, MP, PA, WA \\
\hline
\end{tabular}
\label{tab:paraObs}
\end{table*}

Five epochs of phase--referenced observations~(2013~March, June, August, November and~2015~March) were made of the~6.7--GHz methanol masers in G\,305.200$+$0.019, G\,305.202$+$0.208 and G\,305.208$+$0.206, as well as of the associated quasars J\,1254$-$6111, J\,1256$-$6449 and J\,1312$-$6035 using the LBA. The~J\,2000 coordinates which we used for the maser observations were obtained from the~MMB \citep{Green+12a} for G\,305.200$+$0.019 at $\alpha =$13$^{h}$11$^{m}$16\fs 93, $\delta =-$62$^{\circ}$45\arcmin 55\farcs 1, G\,305.202$+$0.208 at~$\alpha =$~13$^{h}$11$^{m}$10\fs 49, $\delta = -$62$^{\circ}$34\arcmin 38\farcs 8 and G\,305.208$+$0.206 at~$\alpha =$~13$^{h}$11$^{m}$13\fs 71,~$\delta =-$62$^{\circ }$34\arcmin 41\farcs 4. The latter two sources are separated by~22\arcsec \/ and we pointed the telescopes midway between the two source positions for the observations, but correlated the individual sources at the respective~MMB~positions. The primary beams of the individual telescopes are greater than~1\arcmin \/ at~6.7--GHz, and so there is little loss in sensitivity in employing this procedure. Table~\ref{tab:lba-coords} contains updated coordinates for these sources, based on the absolute position of J\,1254$-$6111 (see Section~\ref{sec:data-reduction}). The coordinates for J\,1254$-$6111 and J\,1256$-$6449 in Table~\ref{tab:lba-coords} are from \citet{Petrov+11} with positional uncertainties of~1.49~mas and~1.26~mas respectively. The coordinates which were used for J\,1312$-$6035 were~$\alpha =$~13$^{h}$12$^{m}$12\fs 34, $\delta =-$60$^{\circ}$~35\arcmin ~38\farcs 1, and are accurate to better than~1\arcsec \/ \citep{Murphy+10a}. We present updated coordinates
for this source in Table~\ref{tab:lba-coords}, and used these in our analyses.

Observations typically lasted for~$\lesssim $24~hours~(Table~\ref{tab:paraObs}), with approximately one third of the time used for observations of the~6.7--GHz methanol masers and associated background quasars. The phase--referencing technique involved alternating scans for~2~minutes on the target maser with scans lasting~2~minutes on the nearby~($\sim $2$^\circ $) quasars. We scheduled our observations to ensure that J\,1254$-$6111, J\,1256$-$6449 and J\,1312$-$6035 were phase--referenced to all maser sources.

The phase--referenced observations were interspersed with tropospheric calibration observations which consisted of short~(2~minutes per source) scans of~12~--~18~quasars from the International Celestial Reference Frame~(ICRF) Second Realization catalogue \citep{Ma+09}. The ICRF sources were observed over as broad an azimuth range (generally at low elevation) and were arranged into~45~minute blocks with intervals of between~3 to~6~hours between consecutive blocks.

The data was correlated at Curtin University using the DiFX\footnote{This work made use of the Swinburne University of Technology software correlator, developed as part of the Australian Major National Research Facilities Programme and operated under licence.} software \citep{Deller+11}. We correlated a~2~MHz zoom--band with~2048~channels for the maser data, giving spectral channel width of~0.977~kHz and corresponding velocity separation of~0.055~\kms . Only a fraction of the recorded~32~MHz was used because the maser emission covers a small bandwidth. In contrast we correlated the full recorded bandwidth for the phase  quasar observations. In the~2013~March epoch,~256~spectral channels were used per~16~MHz bandwidth. In the remaining epochs,~32~spectral channels were used per~16~MHz bandwidth corresponding to resolutions of~62.5~kHz and~500~kHz respectively. We used identical correlation parameters for the ICRF and background quasar data for each epoch.

The full details of the observations including the~LBA setup, correlation parameters and calibration procedures can be found in \citet{Krishnan+15}.

\begin{table*}
\caption{Coordinates of the observed sources between~2013~--~2015~March. The separation and position angle columns describe the offset in the sky between the respective quasar and the~6.7--GHz methanol masers in G\,305.200$+$0.019. The offset and separation of the methanol masers in G\,305.202$+$0.208 and G\,305.208$+$0.206 with respect to G\,305.200$+$0.019 is also listed. The reported positions for G\,305.200$+$0.019, G\,305.202$+$0.208, G\,305.208$+$0.206 and J\,1312$-$6035 are revised based on the~2013~June epoch (see Section~\ref{sec:data-reduction}).}
\begin{tabular}{rrcrrrr}
\hline
&     \multicolumn{1}{c}{\bf Source}    &\multicolumn{1}{c}{\bf Separation}  & \multicolumn{1}{c}{\bf Position} & \multicolumn{1}{c}{\bf RA}      & \multicolumn{1}{c}{\bf Dec}  \\
&  \multicolumn{1}{c}{\bf name} & & \multicolumn{1}{c}{\bf angle} &  & \\
& &                 \multicolumn{1}{c}{\bf ($^{\circ}$)}   & \multicolumn{1}{c}{\bf ($^{\circ}$)} & \multicolumn{1}{c}{\bf ($^{h~m~s}$)} & \multicolumn{1}{c}{\bf ($^{\circ}$~\arcmin ~\arcsec)} \\
\hline
\hline
Masers:                          &                     & & & & & \\
                                 & G\,305.200$+$0.019  & --   & -- & 13 11 16.8912 & $-$62 45 55.008 \\
                                 & G\,305.202$+$0.208  & 0.19 & 356.24 & 13 11 10.4904 & $-$62 34 38.856 \\
                                 & G\,305.208$+$0.206  & 0.19 & 358.12 & 13 11 13.7017 & $-$62 34 41.397 \\
Detected quasars:                &                     & & & & & \\
                                 & J\,1254$-$6111      & 2.50 & 307.21 & 12 54 46.5768 & $-$61 11 34.969 \\
                                 & J\,1256$-$6449      & 2.65 & 217.58 & 12 56 03.4030 & $-$64 49 14.817 \\
                                 & J\,1312$-$6035      & 2.17 & 2.99 & 13 12 12.2928 & $-$60 35 38.220 \\
Non--detected quasar:            &                     & & & & & \\
                                 & J\,1259$-$6519      & 2.88 & 205.54 & 12 59 23.9000 & $-$65 19 53.200 \\
\hline
\end{tabular}
\label{tab:lba-coords}
\end{table*}

\section{Data calibration}
\label{sec:data-reduction}

\begin{figure}
  \centering
  \includegraphics[width=\linewidth]{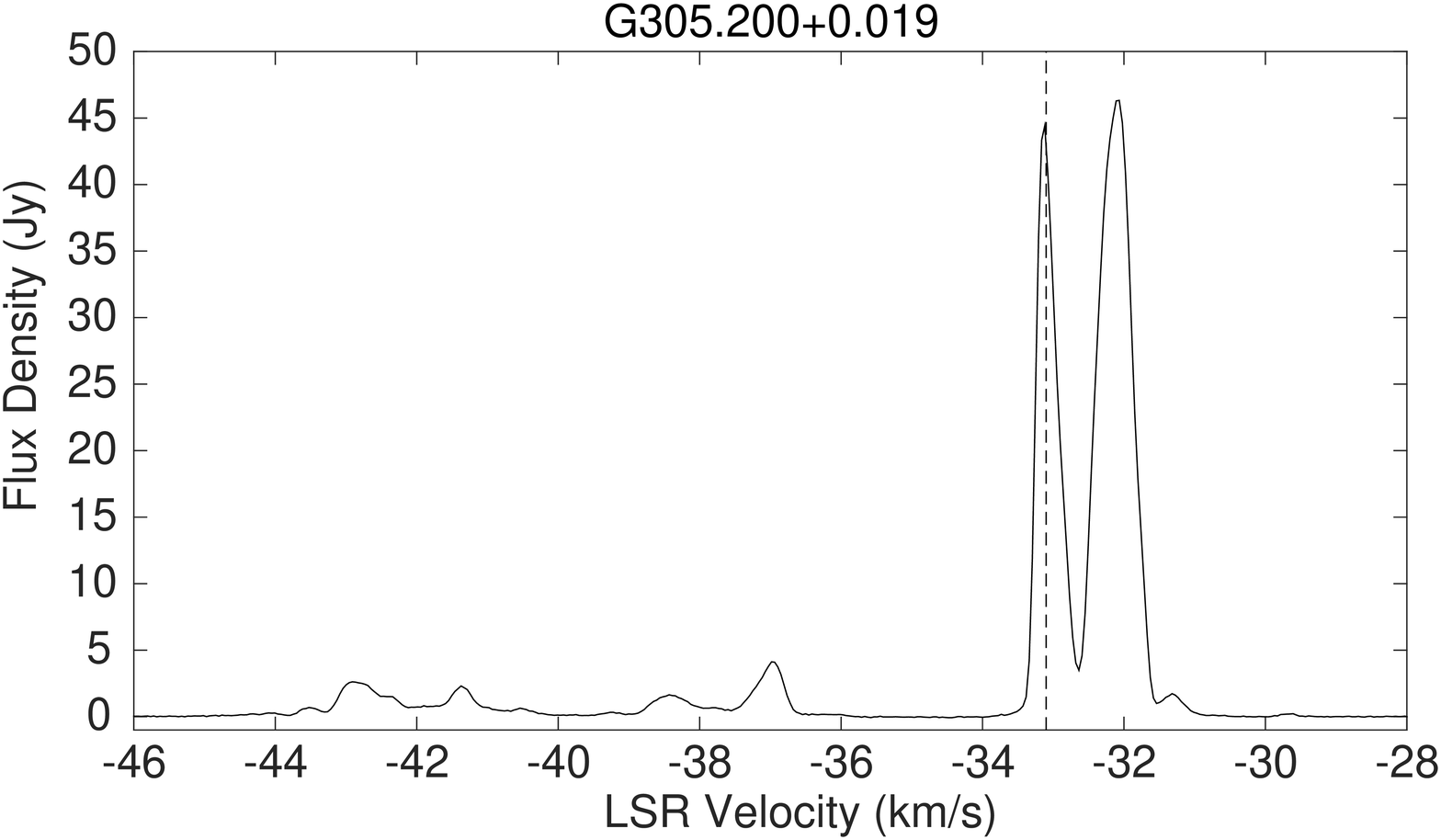}
  \includegraphics[width=\linewidth]{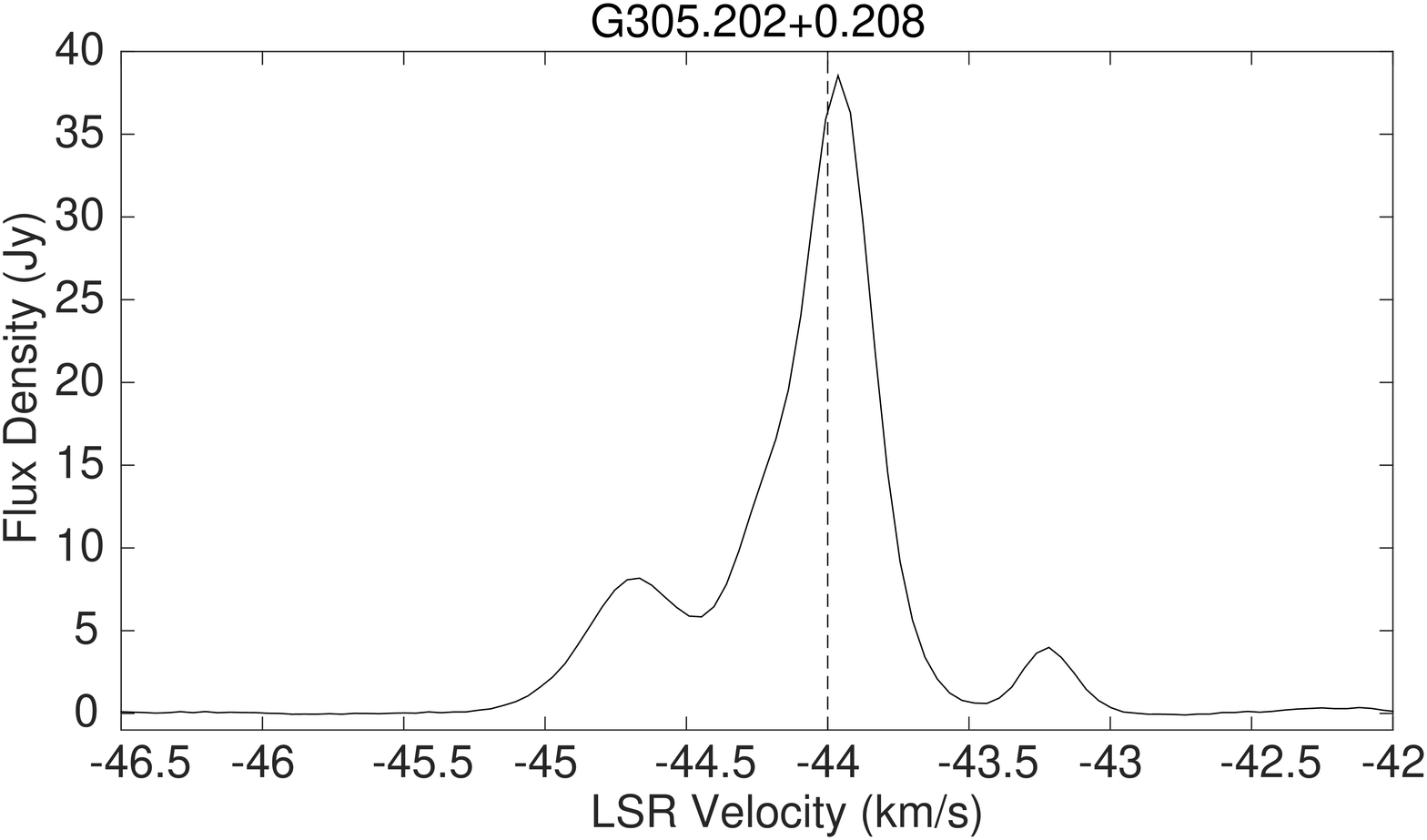}
  \includegraphics[width=\linewidth]{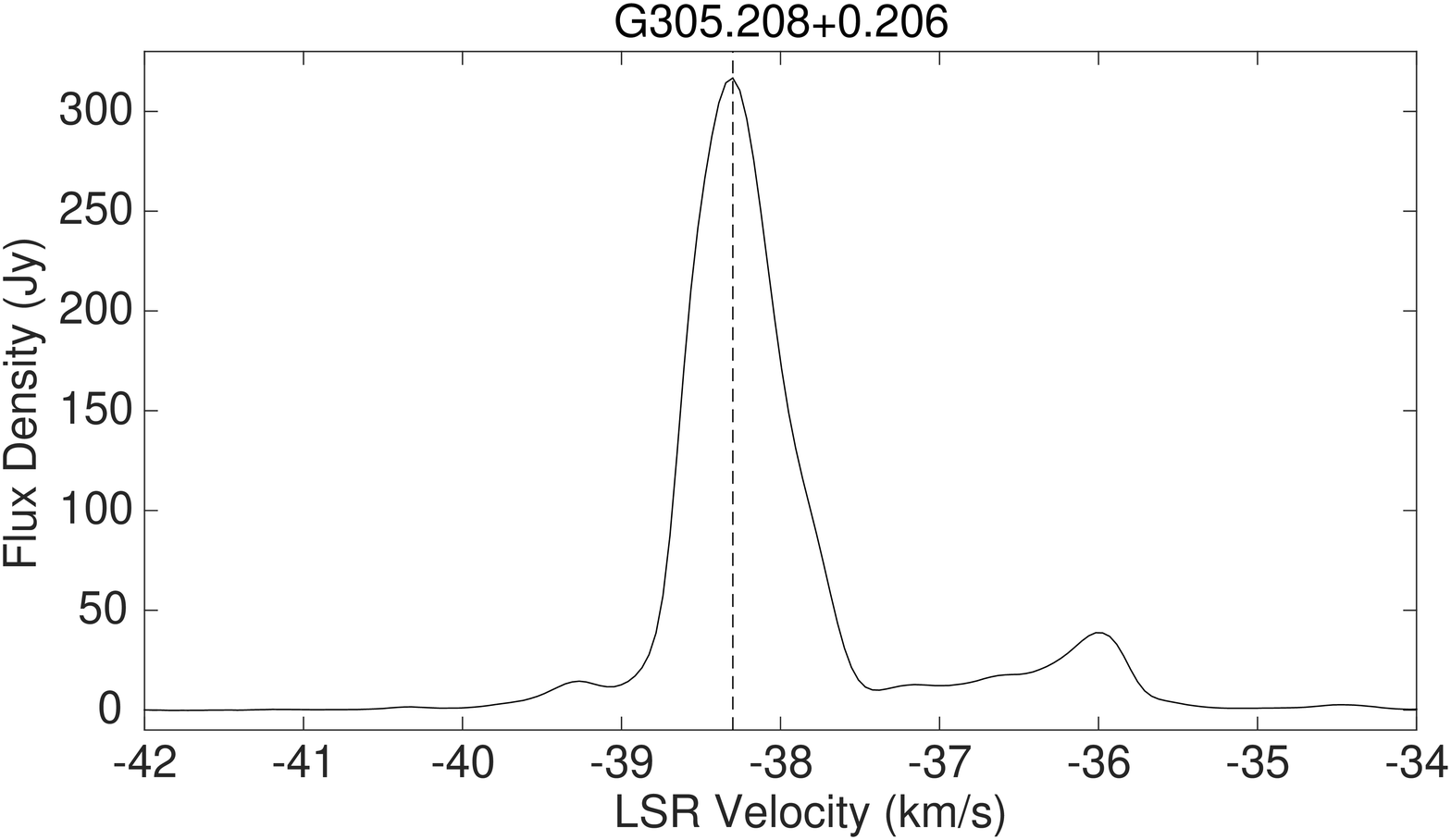}
\caption{The autocorrelation spectra using all antennas from the~2013~November session. \emph{top}: G\,305.199$+$0.005~($-$45.0~to~$-$40.0~\kms ) and G\,305.200$+$0.019~($-$38.0~to~$-$29.5~\kms ); the emission in the region with peak at~$-$38.5~\kms \/ was not not detected in \citet{Green+12a} (see Section~\ref{sec:g-305.200+0.019}), \emph{middle}: G\,305.202$+$0.208 and \emph{bottom}: G\,305.208$+$0.206. The dashed vertical lines indicates the channel which we used for astrometry.}
  \label{fig:G305_200_spec}
\end{figure}

We followed the standard data reduction pathway for VLBI data calibration of the ICRF and phase--referencing mode observations using the Astronomical Image Processing System~\citep[AIPS;][]{Greisen+03}.

We removed the estimated ionospheric delay \citep[determined from global models based on GPS total electron content~(TEC) observations;][]{Walker+99}, the Earth Orientation Parameters~(EOPs), parallactic angle effects and clock drift for each observatory from the ICRF multiband delays before correcting for Doppler effects which manifest as a spectral frequency shift. Figure~\ref{fig:G305_200_spec} shows the autocorrelation spectra of the~6.7--GHz methanol maser emission associated with G\,305.199$+$0.005, G\,305.200$+$0.019, G\,305.202$+$0.208 and G\,305.208$+$0.206. The local standard of rest velocities~($v_{\mbox{\scriptsize lsr}}$) of~$-$42.8,~$-$33.1,~$-$44.0 and~$-$38.3~\kms \/ correspond to the maser spectral peaks of the positions reported by \citet{Green+12a}. We performed amplitude calibration using ACCOR to correct imperfect sampler statistics during correlation and ACFIT to scale the spectra at all observing stations from a single autocorrelation scan of each maser source.

Delay calibration was performed on the phase--reference quasar dataset using J\,1254$-$6111, which was the brightest quasar with the most accurate initial position. We applied the delay solutions (with zeroed rates) to the other phase--reference quasars and employed the procedure in \citet{Krishnan+15} to apply the corrections to the maser dataset.

Due to the relative strength of the maser emission compared to the quasars~(see Tables~\ref{tab:j1254-epochs-II} and~\ref{tab:j1254-epochs}), we treated the masers as the calibrator and the quasar as the target for phase--referencing. In order to select the best maser features for astrometry, we studied the cross correlation spectra and identified the maser features which showed the least flux variations across all baselines and epochs. These were found to be the~$-$33.1~\kms \/ peak for G\,305.200$+$0.019, the~$-$44.0~\kms \/ peak for G\,305.202$+$0.208 and the~$-$38.3~\kms \/ peak for G\,305.208$+$0.206 (Figure~\ref{fig:G305_200_spec}). We then used the associated maser spectral channel for phase calibration, and produced images of the emission, finding them to be point--like and persistent across all epochs. This demonstrated that they were suitable for astrometry, and we transferred the phase solutions to J\,1254$-$6111, J\,1256$-$6035 and J\,1312$-$6035 to complete the phase--reference procedure.

\begin{figure}
 \centering
  \includegraphics[width=0.34\textwidth]{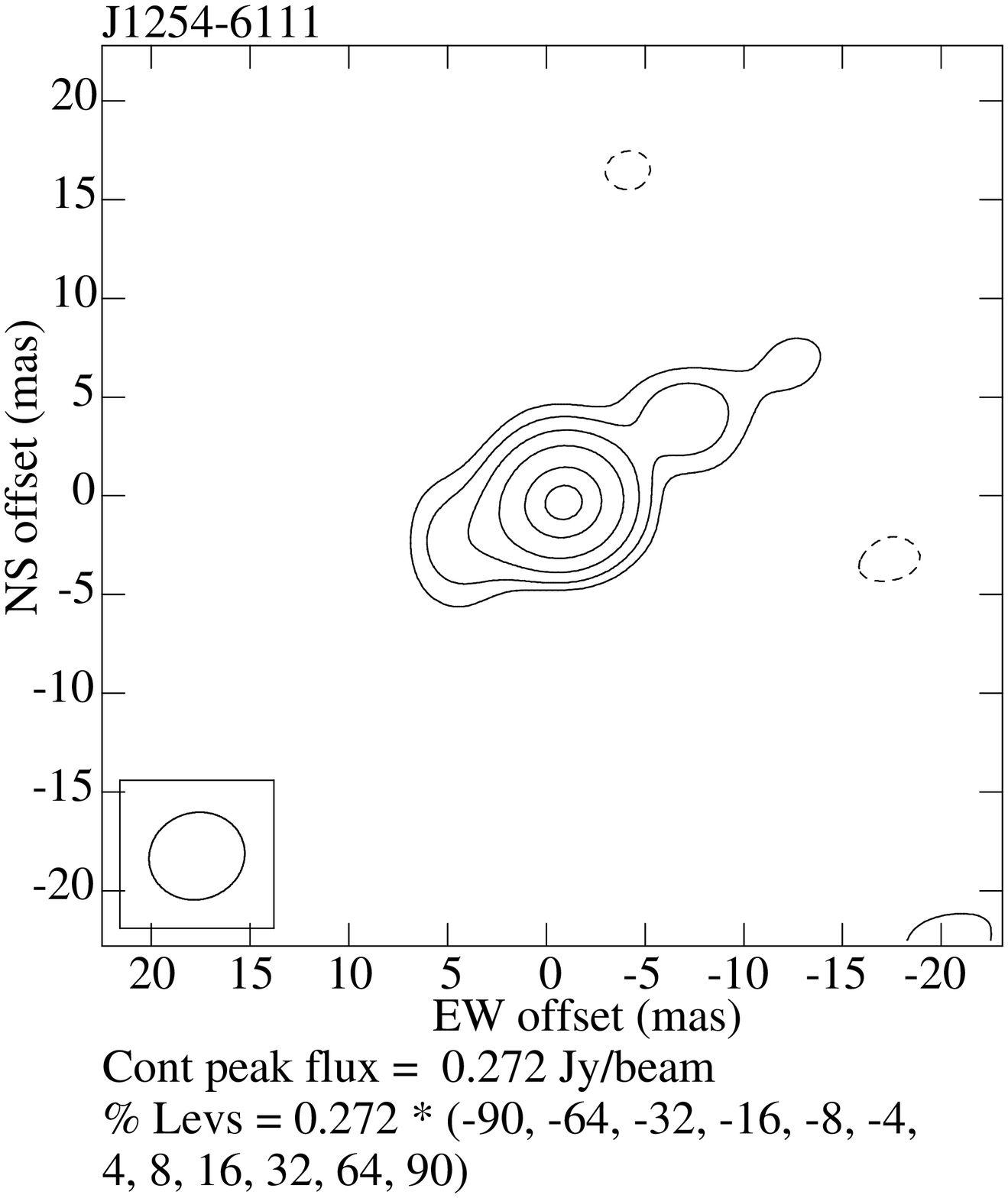}
  \vskip 0.1cm
  \includegraphics[width=0.34\textwidth]{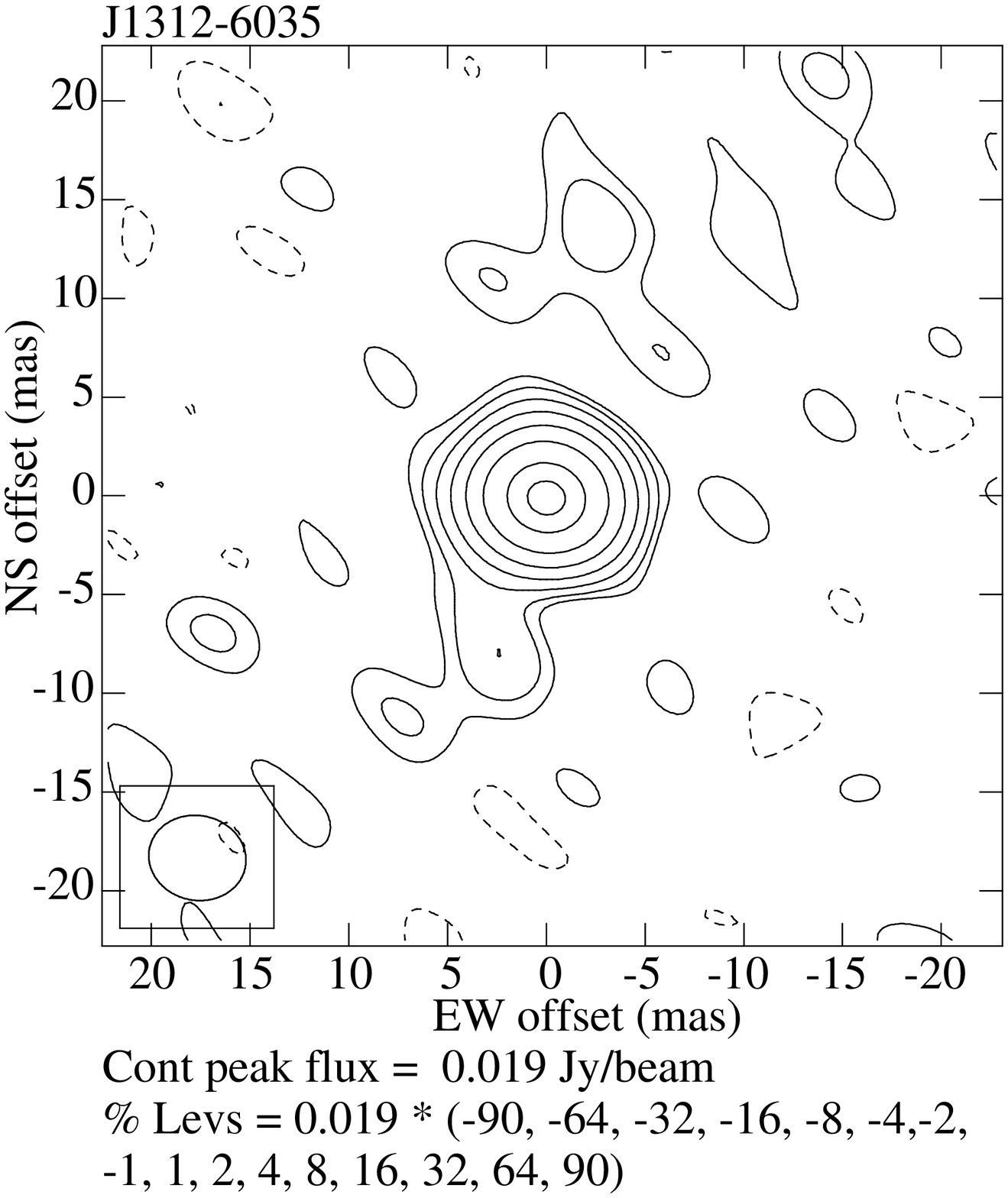}
  \vskip 0.1cm
  \includegraphics[width=0.34\textwidth]{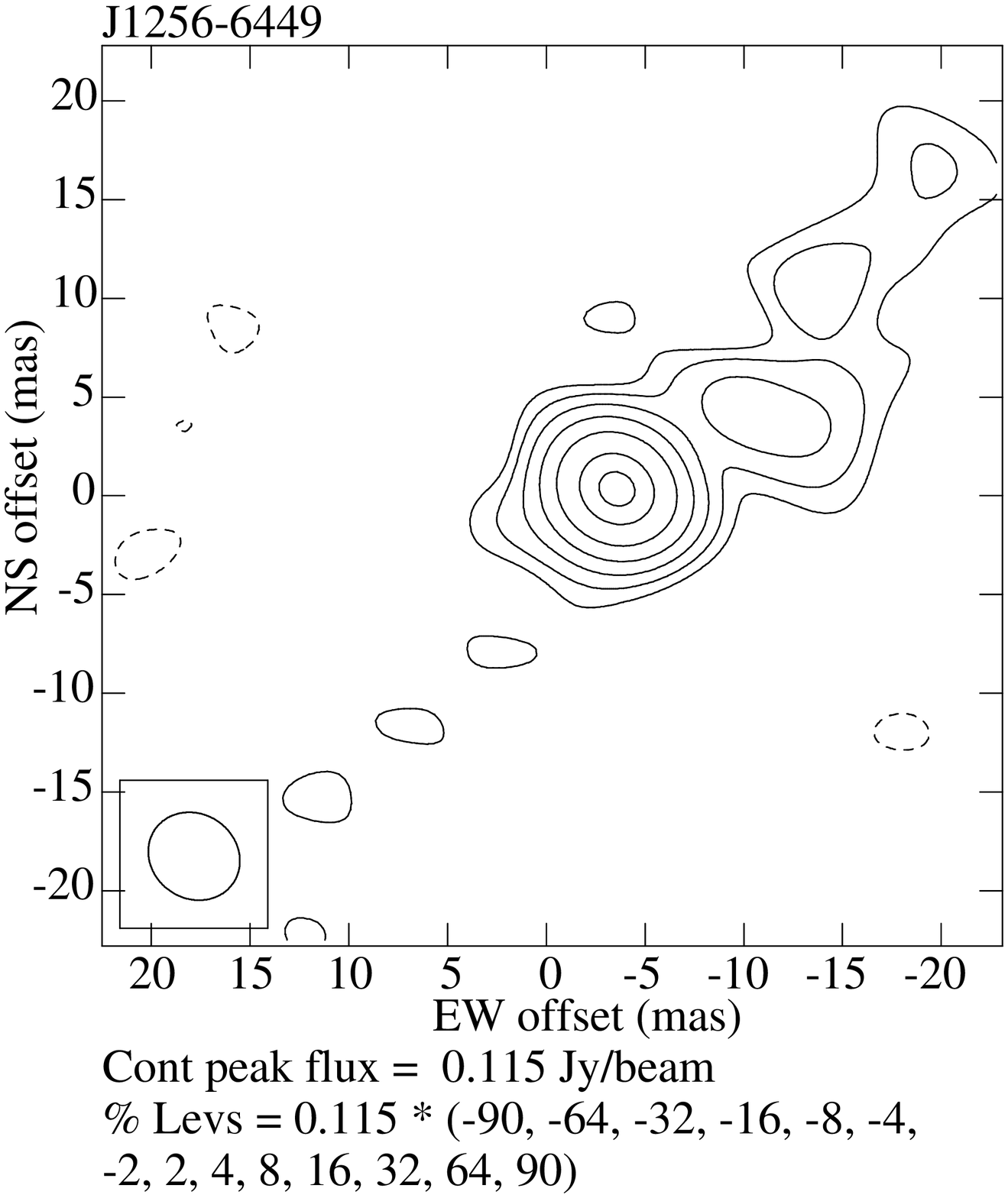}
\caption{J\,1254$-$6111~(\emph{top}), J\,1312$-$6035~(\emph{middle}) and J\,1256$-$6035~(\emph{bottom}) phase--referenced to G\,305.202$+$0.208 from the~2013~June session. The quasars showed consistent centroid structure dominated by a single peak throughout all epochs. There was some variability in the quality of the images from one epoch to another. with image distortions resulting from residual phase errors. However these do not prevent us from accurately measuring the position of the core.}
 \label{fig:G305-quasars}
\end{figure}

We averaged all channels for each phase--reference quasar and produced images of the emission using a Gaussian restoring beam of~$\sim $4.5~mas$^2$~(averaged over all sources and epochs) and report detections for J\,1254$-$6111, J\,1256$-$6035 and J\,1312$-$6035 on VLBI baselines. We also observed J\,1259$-$6519 \citep{McConnell+12} for~6~mins to test its suitability as a quasar for phase--referencing in the~2013~March epoch. We failed to detect it at an upper limit for detection at~5~times the image~RMS (from a box of size~1.5~arcsec$^2$) and hence excluded it from subsequent observations. The detected quasars appear to be dominated by single components, showing deviation from point--like structure at levels~$<$10\% of the peak flux density (Figure~\ref{fig:G305-quasars}). Residual phase errors contribute to distortions in the final images. We located the centroid position of the quasar by fitting a~2\,D~Gaussian to the deconvolved quasar emission using JMFIT and recorded the offset of the emission peak from the centre of the image field for all epochs. We then reversed the signs of the offsets in order to represent the shift of the maser emission with respect to the quasar. These offsets are presented in Tables~\ref{tab:j1254-epochs-II} and~\ref{tab:j1254-epochs} for G\,305.200$+$0.019 and G\,305.202$+$0.208 respectively. The errors in the fitted positions for the offset positions in Tables~\ref{tab:j1254-epochs-II} and~\ref{tab:j1254-epochs} are from JMFIT, and are in agreement with theoretical predictions of astrometric accuracy to within~$\sim $0.02~mas.

The input source position errors are~$\sim $0.4\arcsec \/ for the MMB coordinates for all the maser sources and better than~2~mas for two of the phase--reference quasars. Hence, for these quasars, any offset of the image from the centre of the field is due to the offset of the correlated maser position from the true maser position. We refined the coordinates for the maser feature used for phase--referencing by iteratively adjusting them for the~2013~June epoch. We repeated this until the J\,1254$-$6111 emission was at the centre of the image, post phase--referencing to the maser. We chose J\,1254$-$6111 for this process as it was the brightest of the quasars with the best determined position. We then used the updated maser coordinates for all other epochs. As the phase solutions are derived assuming that the maser is at a given position, accurate coordinates are essential as positional errors produce residual phase shifts which cannot be perfectly modelled as a position shift alone when applied to the quasar. This will result in severe image degradation in the weak quasar sources \citep{Reid+09a,Beasley+95}. From our phase--referenced observations we present updated coordinates for G\,305.200$+$0.019, G\,305.202$+$0.208 and G\,305.208$+$0.206 in Table~\ref{tab:lba-coords}. The offset between the source coordinates in Table~\ref{tab:lba-coords} and \citet{Green+12a} for G\,305.200$+$0.019 is~0.282\arcsec , G\,305.202$+$0.208 is~0.057\arcsec \/ and G\,305.208$+$0.206 is~0.076\arcsec . We also present updated coordinates for J\,1312$-$6035 to better than~2~mas in Table~\ref{tab:lba-coords}, with an offset of~0.368\arcsec \/ from the \citet{Murphy+10a} position.

The change in the position of the maser feature used for astrometry was modelled independently in right ascension and declination and included corrections for the ellipticity of Earth's orbit \citep{Reid+14a}. We assigned \emph{a priori} astrometric uncertainties in right ascension and declination to account for systematic uncertainties in each coordinate. These estimates were iteratively adjusted until the~$\chi ^2 _\nu $ value per degree of freedom of~$\sim $1 was obtained for the parallax model. These uncertainties were added in quadrature to the formal errors of the offsets in Tables~\ref{tab:j1254-epochs-II} and~\ref{tab:j1254-epochs} to obtain the parallax which we present in the following section.

We have not included the data obtained from the Hartebeesthoek~26m and Warkworth~30m antennas for the results presented in this paper. The location of the Hartebeesthoek antenna with respect to the rest of the LBA affects its participation in the ICRF observations, preventing us from correcting the clock rate at this observatory. We are currently exploring alternative methods to determine the clock rate from Hartebeesthoek. The Warkworth antenna was included in the array on a test basis for the first time for maser astrometry in~2015~March (see Table~\ref{tab:paraObs}). However, we were unable to obtain sufficient valid data from this antenna using our current calibration pipeline. We anticipate that astrometric results from measurements including Warkworth will be presented in the future as these issues are resolved.

\section{Parallax measurements}
\label{sec:results-G305-20}

\begin{table*}
\caption{The differential fitted positions of the~$-$33.1~\kms\/ feature in G\,305.200$+$0.019 with respect to J\,1254$-$6111, J\,1312$-$6035 and J\,1256$-$6449 and the corresponding parallax. The flux density of the maser peak emission corresponding to the~$-$33.1~\kms\/ feature is also presented. We used an emission free channel to obtain the maser RMS. The listed RMS for all images have been obtained using the histogram in IMEAN.}
  \centering
\begin{tabular}{rrlrrrccccr}
\hline
           &          &                   &           & &           & \multicolumn{2}{l}{\bf Quasar}& \multicolumn{2}{l}{\bf G\,305.200$+$0.019}&                               \\
\bf Source &\bf Epoch & \bf x offset      & \bf Error & \bf y offset & \bf Error & \bf flux density &  \bf RMS   & \bf flux density &  \bf RMS  & \bf Parallax                  \\
           &          & \multicolumn{2}{c}{\bf (mas)} & \multicolumn{2}{c}{\bf (mas)} & \multicolumn{2}{c}{\bf (mJy)} & \multicolumn{2}{c}{\bf (Jy)} & \multicolumn{1}{c}{\bf (mas)} \\
\hline
\hline
 J\,1254$-$6111 & 2013.210 &    2.362  & 0.056 & $-$0.372 & 0.032 & 172.6 & 1.7 & 28.34 & 0.02 & 0.20$\pm $0.08\\
                & 2013.460 &    0.381  & 0.007 & $-$0.235 & 0.009 & 302.3 & 1.2 & 28.56 & 0.02 & \\
                & 2013.621 & $-$0.793  & 0.023 & $-$0.136 & 0.017 & 159.7 & 1.2 & 28.00 & 0.02 & \\
                & 2013.887 & $-$2.335  & 0.021 &    0.001 & 0.021 & 159.2 & 3.6 & 25.46 & 0.02 & \\
                & 2015.236 & $-$11.479 & 0.020 & $-$0.665 & 0.014 & 183.1 & 1.2 & 28.43 & 0.03 & \\
 \\
 J\,1312$-$6035 & 2013.210 &    1.398  & 0.036 &    0.424 & 0.018 & 18.8  & 2.3 &   &   & 0.21$\pm $0.09\\
                & 2013.460 & $-$0.533  & 0.031 &    0.694 & 0.040 & 14.8  & 2.7 &   &   & \\
                & 2013.621 & $-$1.789  & 0.023 & $-$0.317 & 0.019 & 17.5  & 1.8 &   &   & \\
                & 2013.887 & $-$3.228  & 0.103 & $-$0.981 & 0.099 & 5.7   & 2.3 &   &   & \\
                & 2015.236 & $-$12.066 & 0.041 & $-$1.327 & 0.029 & 17.3  & 2.5 &   &   & \\
 \\
 J\,1256$-$6449 & 2013.210 &    3.988  & 0.011 & $-$1.808 & 0.009 & 87.7  & 1.2 &   &   & 0.40$\pm $0.33\\
                & 2013.460 &    3.012  & 0.018 & $-$0.564 & 0.022 & 81.4  & 8.6 &   &   & \\
                & 2013.621 &    2.359  & 0.039 & $-$0.987 & 0.026 & 56.9  & 8.8 &   &   & \\
                & 2013.887 &    1.815  & 0.071 & $-$1.035 & 0.066 & 26.4  & 1.5 &   &   & \\
                & 2015.236 & $-$9.062  & 0.012 & $-$1.292 & 0.009 & 63.6  & 3.1 &   &   & \\
\hline
\hline
\end{tabular}
\label{tab:j1254-epochs-II}
\end{table*}

\begin{figure*}
 \centering
 \includegraphics[width=\linewidth]{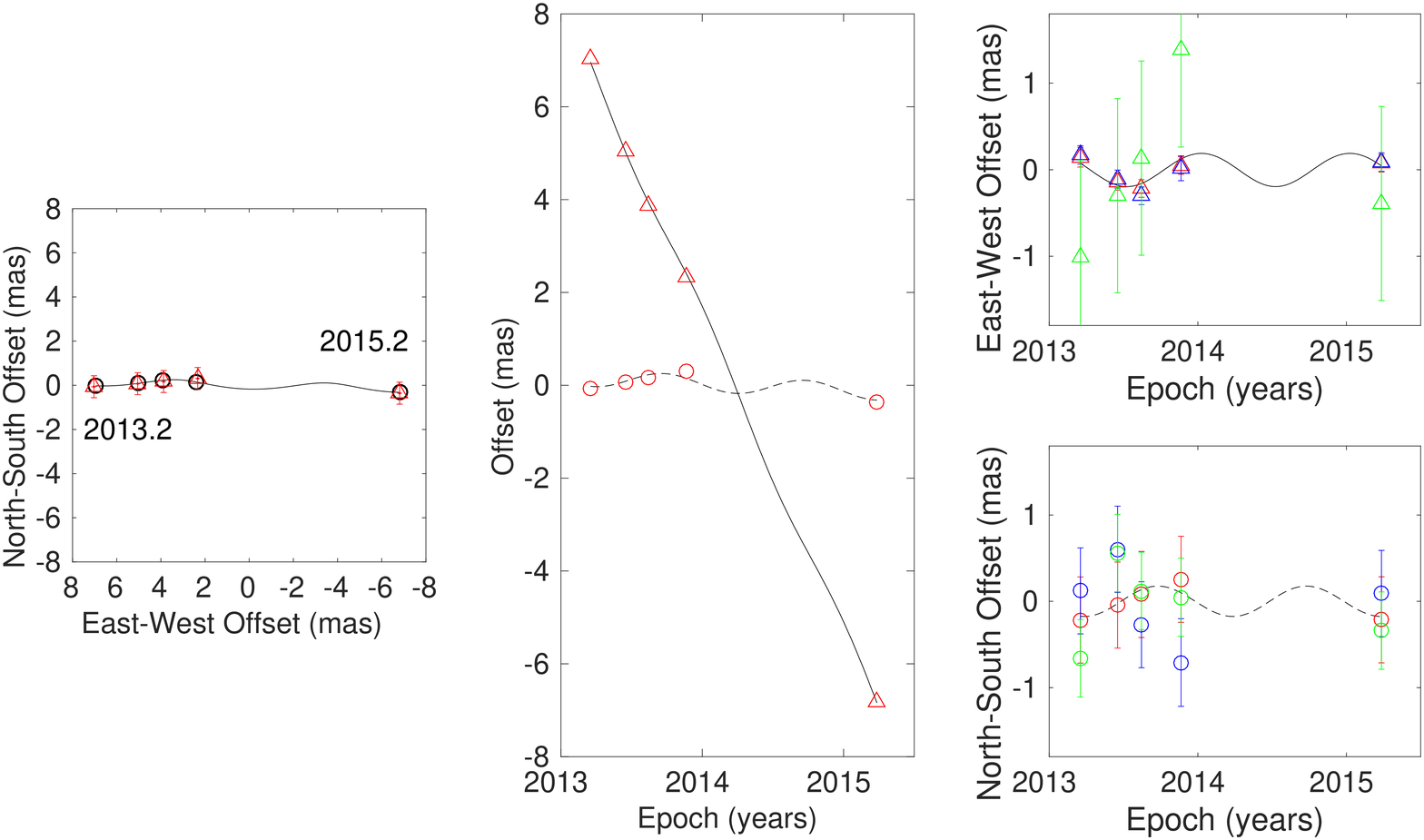}
\caption{Parallax and proper motion of the $-$33.1~\kms\/ reference feature in G\,305.200$+$0.019 with respect to J\,1254$-$6111~(red), J\,1312$-$6035~(blue) and J\,1256$-$6449~(green). Left panel: the sky positions with the first and last epochs labeled. We have opted to show the positions with respect to a single quasar for clarity. The expected positions from the fits are indicated with black circular markers. Middle panel: east--west (triangles) and north--south (circles) motion of the position offsets and best combined parallax and proper motions fits versus time. Right panels: the east--west~(top) and north--south~(bottom) parallax signature with the best fit proper motions removed.}
\label{fig:G305_200_para}
\end{figure*}

\begin{figure*}
 \centering
 \includegraphics[width=\linewidth]{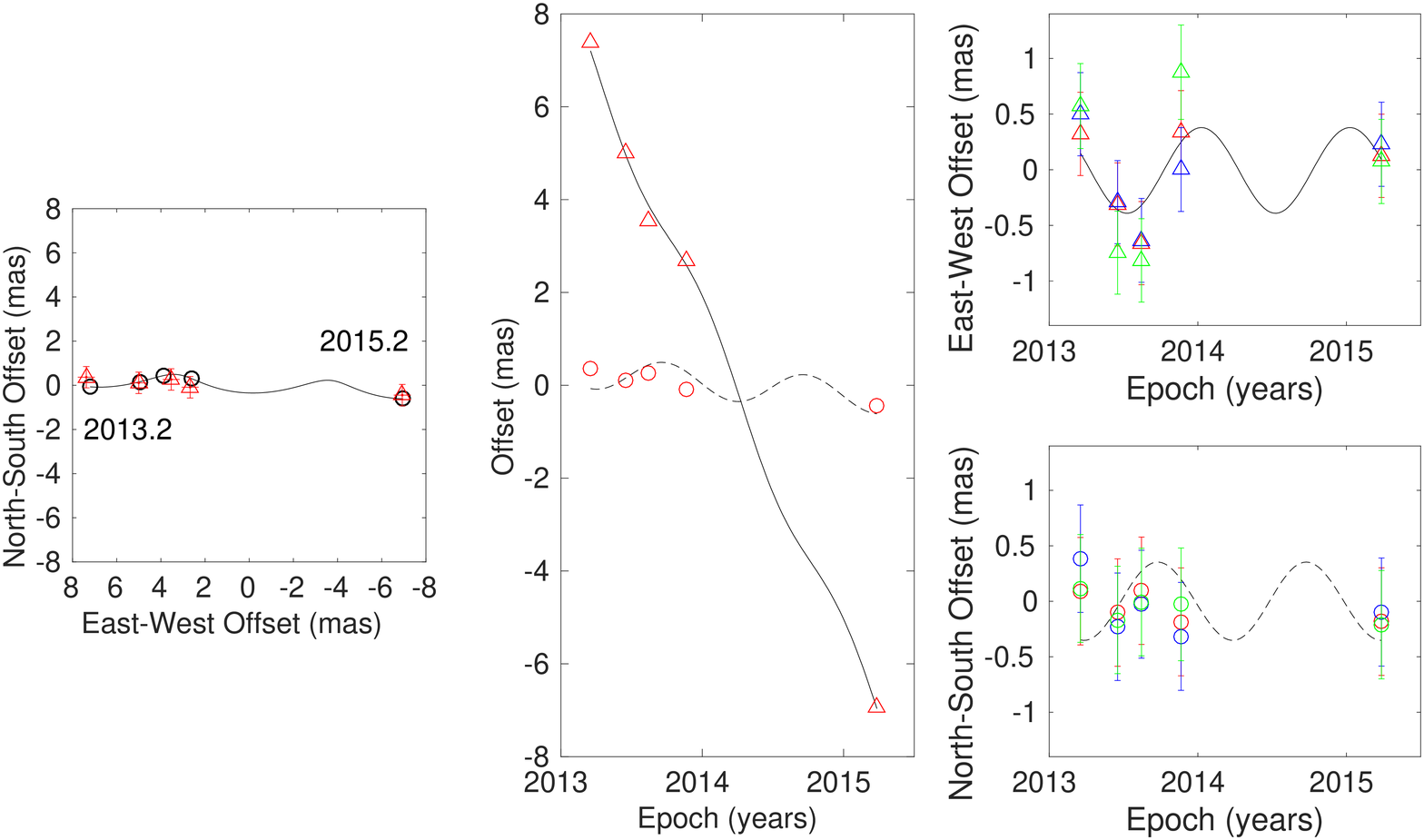}
\caption{Parallax and proper motion of the $-$44.0~\kms\/ reference feature in G\,305.202$+$0.208 with respect to J\,1254$-$6111~(red), J\,1312$-$6035~(blue) and J\,1256$-$6449~(green). Left panel: the sky positions with the first and last epochs labeled. We have opted to show the positions with respect to a single quasar for clarity. The expected positions from the fits are indicated with black circular markers. Middle panel: east--west~(triangles) and north--south~(circles) motion of the position offsets and best combined parallax and proper motions fits versus time. Right panels: the east--west~(top) and north--south~(bottom) parallax signature with the best fit proper motions removed.}
\label{fig:G305_20_para}
\end{figure*}

The parallax of G\,305.200$+$0.019 is measured to be~0.21$\pm $0.06~mas (Table~\ref{tab:j1254-epochs-II} and Figure~\ref{fig:G305_200_para}) and is a variance weighted average derived from measurements with respect to J\,1254$-$6111, J\,1312$-$6035 and J\,1256$-$6449. The variance weighted average parallax of G\,305.202$+$0.208 is measured to be~0.42$\pm $0.13~mas (Table~\ref{tab:j1254-epochs} and Figure~\ref{fig:G305_20_para}) with respect to J\,1254$-$6111, J\,1312$-$6035 and J\,1256$-$6449. The sources are almost certainly contained within the same Giant Molecular Cloud~(GMC), and so we have averaged the values to present an estimated parallax to the~G\,305.2 region in general. Due to the large uncertainty in the G\,305.202$+$0.208 measurement, we have adopted a variance weighted approach to obtain a parallax of~0.25$\pm $0.05~mas, corresponding to~4.1$^{+1.2}_{-0.7}$~kpc, which we present as the current best estimate of the distance to the~G\,305.2 region in Table~\ref{tab:parResII}.

In order to constrain errors in the measured proper motions, we made image cubes of the maser emission and analyzed the changes in the spot distribution from~2013~to~2015~March. We added the median of the spot motion distribution to the modelled proper motion~$(\mu _x , \mu _y )$ of the~$-$33.1~\kms\/ feature in G\,305.200$+$0.019 and of the~$-$44.0~\kms \/ feature in G\,305.202$+$0.208, and included half of the spread from the respective distributions to the formal model errors. This was done to account for determining the motion of the~HMSFR region from a single maser spot. We report the proper motion of the~$-$33.1~\kms \/ feature in G\,305.200$+$0.019 with errors as~$\mu_{x} = -$6.69$\pm $0.03~mas~yr$^{-1}$ and~$\mu_{y} = -$0.60$\pm $0.14~mas~yr$^{-1}$, corresponding to~$-$130.0 and~$-$11.7~\kms \/ at a distance of~4.1~kpc. The proper motion of the~$-$44.0~\kms \/ feature in G\,305.202$+$0.208 is~$\mu_{x} = -$7.14$\pm $0.17~mas~yr$^{-1}$ and~$\mu_{y} = -$0.44$\pm $0.21~mas~yr$^{-1}$, corresponding to~$-$138.8 and~$-$8.6~\kms \/ at a distance of~4.1~kpc.

The uncertainties in~$(\mu _x , \mu _y )$ for both sources correspond to internal motions of~$\lesssim $10~\kms \/ in the maser emission, and are consistent with proper motion estimates of~6.7--GHz methanol masers in HMSFRs \citep[e.g.][]{Moscadelli+14,Sugiyama+14,Goddi+11}. A more detailed analysis of the internal motions of the~6.7--GHz emission in the G\,305.2 sources is beyond the scope of the current text and will be the subject of future analysis.

We were unable to determine the parallax of G\,305.208$+$0.206 from our observations. It is possible that the observed blending of the emission in contiguous channels of the strongest features at~$-$38~\kms \/ and around~$-$36~\kms \/ (see Section~\ref{sec:g-305.202+0.208-g}) undermines the assumption that the emission is point--like at VLBI resolution, and indicates that this is a source which may be unsuitable for sub--milliarcsecond astrometry for the purposes of parallax determination.

The dominant source of errors in phase--referencing at observations~$>$5--GHz, comes from the clocks and unmodeled troposphere \citep{Mioduszewski+09}. \citet{Krishnan+15} examined how the difficulties in effectively correcting for ionospheric phase can affect the multiband delay solutions from the ICRF observations for the LBA. We found an RMS noise of~$\sim $0.1~nsec in the multiband delays from the ICRF observations, which can correspond to large path lengths of~10s of~cm. We therefore applied the linear clock drift rate from the multiband delays to the phase--referenced data and omitted the zenith atmospheric delay corrections during data calibration. We assess that the main sources of error in our parallax measurements are from atmospheric effects which we are currently working to improve.

\begin{table*}
  \centering
\caption{The differential fitted positions of the $-$44.0~\kms\/ feature in G\,305.202$+$0.208 with respect to J\,1254$-$6111, J\,1256$-$6449 and J\,1312$-$6035 and the corresponding parallax. The flux density of the maser peak emission corresponding to the~$-$44.0~\kms\/ feature is also presented. We used an emission free channel to obtain the maser RMS. The listed RMS for all images have been obtained using the histogram in IMEAN.}
\begin{tabular}{rrlrrrccccr}
\hline
           &          &                   &           & &           & \multicolumn{2}{l}{\bf Quasar}& \multicolumn{2}{l}{\bf G\,305.202$+$0.208}&                                    \\
\bf Source &\bf Epoch & \bf x offset      & \bf Error & \bf y offset & \bf Error & \bf flux density &  \bf RMS   & \bf flux density &  \bf RMS  & \bf Parallax                       \\
           &          & \multicolumn{2}{c}{\bf (mas)} & \multicolumn{2}{c}{\bf (mas)} & \multicolumn{2}{c}{\bf (mJy)} & \multicolumn{2}{c}{\bf (Jy)} & \multicolumn{1}{c}{\bf (mas)} \\
\hline
\hline
J\,1254$-$6111 & 2013.210 & 3.240     & 0.034 & 0.583    & 0.023 & 195.1 & 4.3 & 20.56 & 0.02 &0.36$\pm $0.22 \\
               & 2013.460 & 0.865     & 0.022 & 0.324    & 0.021 & 273.3 & 3.0 & 26.60 & 0.03 &\\
               & 2013.621 & $-$0.604  & 0.024 & 0.478    & 0.021 & 174.7 & 2.7 & 22.08 & 0.02 &\\
               & 2013.887 & $-$1.462  & 0.047 & 0.127    & 0.047 & 236.3 & 6.8 & 19.86 & 0.02 &\\
               & 2015.236 & $-$11.068 & 0.033 & $-$0.229 & 0.030 & 114.5 & 1.7 & 16.31 & 0.07 &\\
\\
J\,1312$-$6035 & 2013.210 & 2.508     & 0.029 & 0.959    & 0.021 & 16.7  & 2.8 &    &   & 0.21$\pm $0.22 \\
               & 2013.460 & $-$0.253  & 0.021 & 0.124    & 0.018 & 19.7  & 1.8 &    &   &\\
               & 2013.621 & $-$1.529  & 0.051 & 0.259    & 0.048 & 11.0  & 2.9 &    &   &\\
               & 2013.887 & $-$2.740  & 0.038 & $-$0.368 & 0.033 & 10.8  & 4.3 &    &   &\\
               & 2015.236 & $-$11.947 & 0.062 & $-$1.507 & 0.059 & 17.4  & 4.7 &    &   &\\
\\
J\,1256$-$6449 & 2013.210 & 6.775     & 0.082 & $-$0.291 & 0.046 & 39.5  & 1.7 &    &   & 0.68$\pm $0.23 \\
               & 2013.460 & 3.611     & 0.018 & $-$0.620 & 0.016 & 128.4 & 1.2 &    &   &\\
               & 2013.621 & 2.352     & 0.040 & $-$0.487 & 0.036 & 54.1  & 1.4 &    &   &\\
               & 2013.887 & 2.078     & 0.200 & $-$0.555 & 0.154 & 30.1  & 2.8 &    &   &\\
               & 2015.236 & $-$8.689  & 0.064 & $-$0.979 & 0.065 & 52.4  & 1.8 &    &   &\\
\hline
\end{tabular}
\label{tab:j1254-epochs}
\end{table*}

\begin{table*}
\caption{Summary of the parallax, distance and proper motion measurements of G\,305.200$+$0.019 and G\,305.202$+$0.208.}
\centering
\begin{tabular}{cccc}\hline
 \bf Source &\bf Parallax            &\bf $\mu _x$        & \bf $\mu _y$ \\
 \bf name   &\bf (mas)               & \bf (mas~yr$^{-1}$) & \bf (mas~yr$^{-1}$) \\
\hline
 G\,305.200$+$0.019 & 0.21$\pm $0.06 & $-$6.69$\pm $0.03  & $-$0.60$\pm $0.14 \\
 G\,305.202$+$0.208 & 0.42$\pm $0.13 & $-$7.14$\pm $0.17  & $-$0.44$\pm $0.21 \\
\hline
\multicolumn{1}{r}{\bf Variance weighted average:}&\multicolumn{1}{c}{0.25$\pm $0.05} & & \\
\multicolumn{1}{r}{\bf Corresponding distance:}&\multicolumn{1}{c}{~~~~~4.1$^{+1.2}_{-0.7}$~kpc} & & \\
\hline
\end{tabular}
\label{tab:parResII}
\end{table*}

\section{Properties of associated high--mass star formation regions}
\label{sec:prop-assoc-high}

\subsection{Peculiar motion}
\label{sec:peculiar-motion}
We used the~$v_{\mbox{\scriptsize lsr}}$ of the associated~$^{13}$CO~molecular clump in Table~\ref{tab:CO-clumps} with the measured proper motions in Table~\ref{tab:parResII} to determine the full~3D~motion of G\,305.200$+$0.019 and G\,305.202$+$0.208 in the Galactic Plane which we present in Table~\ref{tab:G305-pec-motions}. The dynamical model of the Galaxy we use is derived from \citet{Reid+14a} and assumes a flat rotation curve of the disk with a circular rotation speed of~$\Theta _0 = 240$~\kms \/ at the radius of the Sun. The distance of the Sun from the Galactic Centre is taken to be~$R_0 = 8.34$~kpc, and the Solar Motion components are~$U_\odot = 10.70$~\kms \/ (towards the Galactic Centre),~$V_\odot = 15.60$~\kms \/ (clockwise and in the direction of Galactic rotation as viewed from the North Galactic Pole) and~$W_\odot = 8.90$~\kms \/ (in the direction of the North Galactic Pole). In \citet{Reid+14a} there is a good fit to the model of spiral arm motions when an~RMS of about~5~--~7~\kms \/ is assumed for each velocity component of HMSFR, which is reasonable for virial motions of stars in giant molecular clouds. Table~\ref{tab:G305-pec-motions} shows that at a distance of~4.1~kpc the calculated peculiar velocity components for G\,305.200$+$0.019 and G\,305.202$+$0.208 follow this trend inside the limits of uncertainty, though there appear to be deviations from the model for some components.

\begin{table*}
  \caption{Peculiar motions in a reference frame that is rotating with the Galaxy.}
  \centering
  \begin{tabular}{lcccrrr}
\hline
 \multicolumn{1}{c}{\bf Source} & \multicolumn{1}{c}{\bf D}    & \multicolumn{1}{c}{\bf $U$}    & \multicolumn{1}{c}{\bf $V$}     & \multicolumn{1}{c}{\bf $W$}     \\
 \multicolumn{1}{c}{\bf name}   & \multicolumn{1}{c}{\bf (kpc)}& \multicolumn{1}{c}{\bf (\kms )}& \multicolumn{1}{c}{\bf (\kms )} & \multicolumn{1}{c}{\bf (\kms )} \\
\hline
\hline
  G\,305.200$+$0.019  & 4.1 & \,~~0.9$\pm $6.2 & \,~~0.5$\pm $6.2 & 7.2$\pm $3.0 \\
  G\,305.202$+$0.208  & 4.1 &  $-$7.9$\pm $6.1 &  $-$0.2$\pm $7.6 &10.8$\pm $4.4 \\
\hline
  \end{tabular}
  \label{tab:G305-pec-motions}
\end{table*}

\subsection{Spiral arm allocation}
\label{sec:spir-arm-alloc}

We have used Galactic~CO~emission measurements by \citet{Reid+16,Garcia+14} with the parallax distance in Table~\ref{tab:parResII} to determine the spiral arm allocation of the G\,305.2 region. It appears that the region could either belong to the Centaurus or Carina spiral arms. In comparing the~$v_{\mbox{\scriptsize lsr}}$ of G\,305.21$+$0.21~(Table~\ref{tab:CO-clumps}) with Figure~13~of \citet{Reid+16}, the sources would favour the Centaurus spiral arm. However their location at a Galactic longitude of~305$^\circ $ places them a few degrees outside the tangent point of~$\sim $310$^\circ $. In terms of the Carina arm, there is a relatively large disparity of~$\sim $15~\kms \/ in the~$v_{\mbox{\scriptsize lsr}}$.

We attempted to constrain the spiral arm association of G\,305.2 by studying the modelled peculiar motions of G\,305.200$+$0.019 and G\,305.202$+$0.208 over a range of distances between~1 and~7~kpc. Our tests indicate reasonable peculiar motions expected for HMSFRS \citep{Reid+09b} between~1~--~4~kpc~(see Table~\ref{tab:G305-pec-motions}), corresponding to the Carina spiral arm, and for distances~$>$5~kpc corresponding to Centaurus. We concluded that the latter contradicts the parallax distance and exacerbates the longitude problem, and therefore find strong evidence to associate the G\,305.2 region with the Carina--Sagittarius arm. This arm has been modelled by \citet{Sato+14} as a log--periodic spiral with a pitch angle of~$\psi = 19.8 \pm 3.1^\circ $ in the Galactocentric azimuth~($\beta $) range between~$3.3^\circ < \beta < 100.9^\circ $. We used the Bayesian Markov chain Monte Carlo~(McMC) procedure from \citet{Reid+14a} to include the G\,305.2 region to this model. In doing so we increase the range of~$\beta $ by~$\sim $40$^\circ $, to obtain an updated pitch angle of~$\psi = 19.0 \pm 2.6^\circ $ for the Carina--Sagittarius arm, now with greater confidence for the range between~$-34.8^\circ < \beta < 100.9^\circ $.

\citet{Honig+15} show that the pitch angles of the arms of four nearby spirals typically fall within the range of~$10^\circ < \psi < 30^\circ $ and are highly scattered. This appears to be the case for arms in different galaxies, among different arms within a galaxy and also within individual spiral arms. We find our updated value of~$\psi $ to be consistent when we model the arm over~$-40^\circ < \beta < 40^\circ $ and~$-40^\circ < \beta < 120^\circ $. This is consistent with \citet{Honig+15} who observe the stability of~$\psi $ along arm segments over~5~--~10~kpc, which is applicable to our case.

\subsection{Source details}
\label{sec:phys-constr-ioniz}

\begin{figure}
  \centering
  \includegraphics[width=1.1\linewidth ]{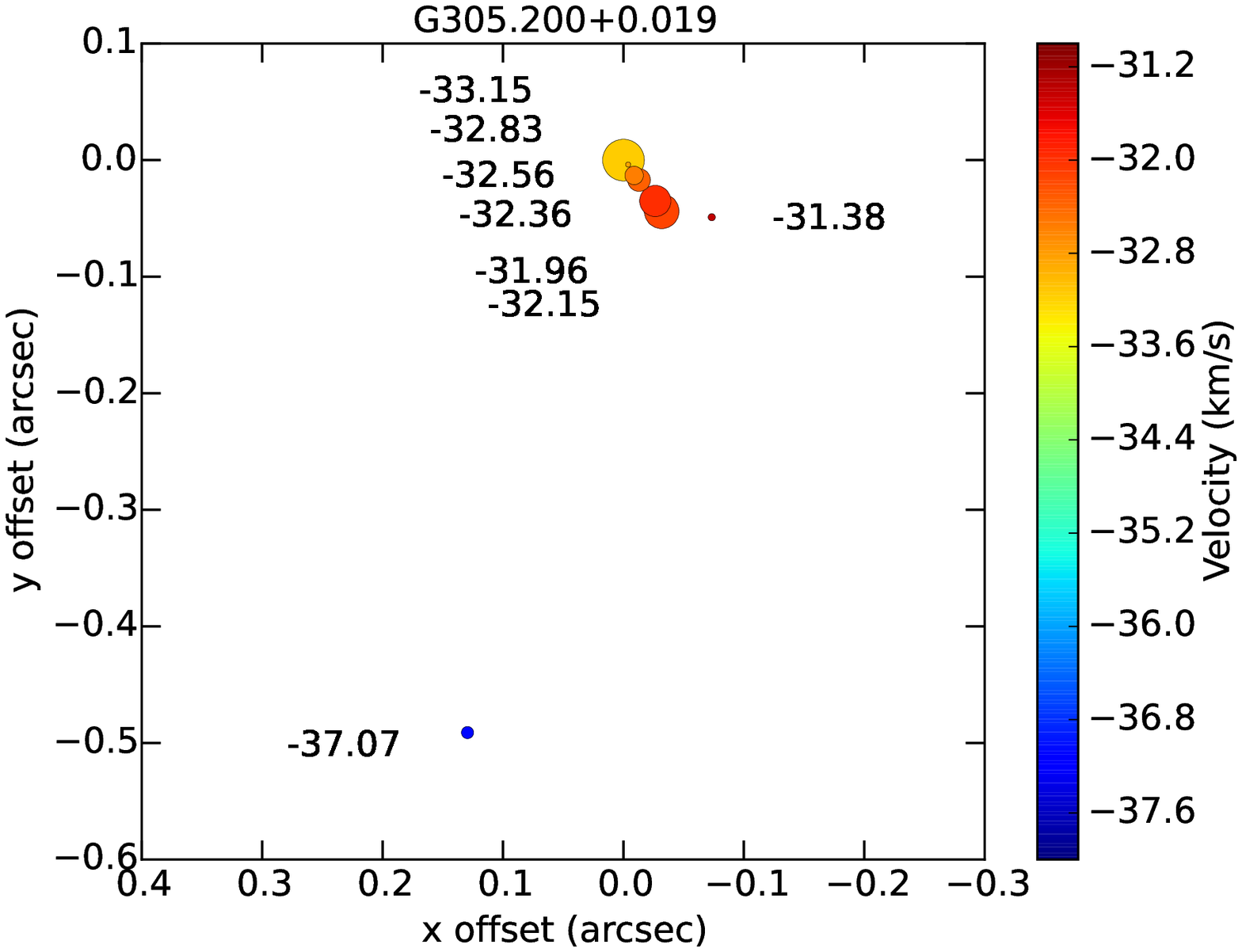}
  \includegraphics[width=1.1\linewidth ]{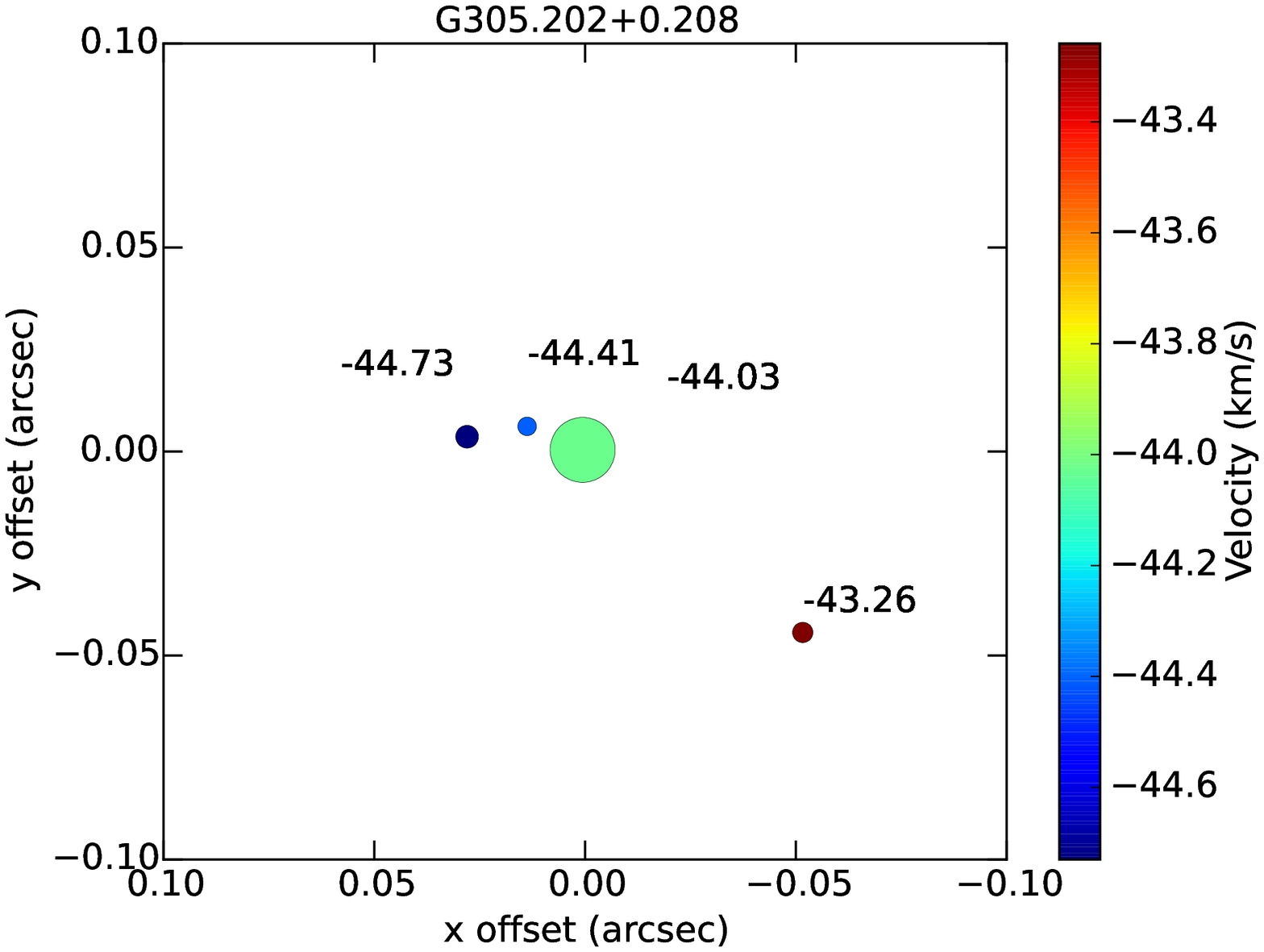}
  \includegraphics[width=1.1\linewidth ]{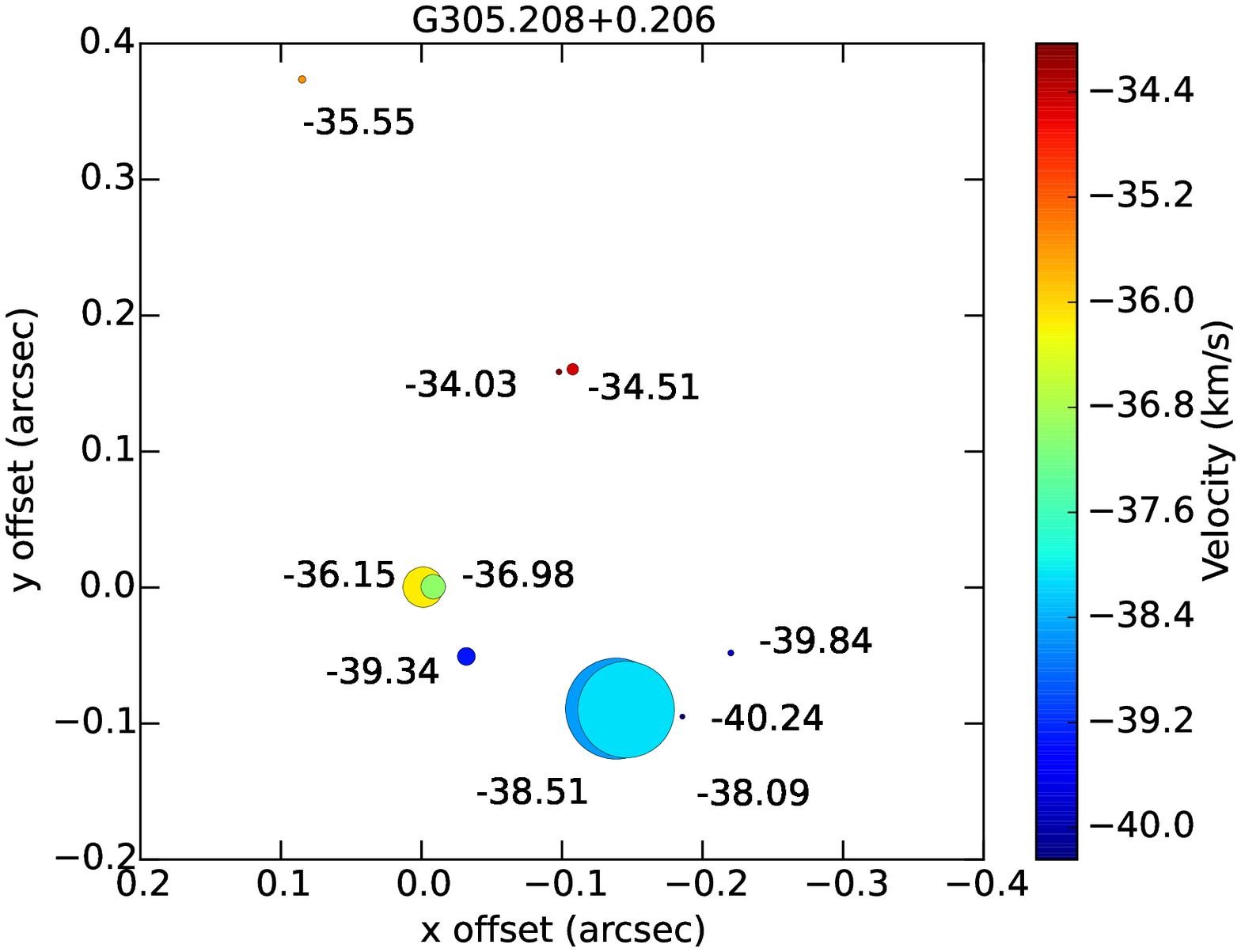}
\caption{The 6.7--GHz methanol maser emission for G\,305.200$+$0.019~(\emph{top}), G\,305.202$+$0.208~(\emph{middle}) and G\,305.208$+$0.206~(\emph{bottom}) from the~2013~November session. The numbers in the image field correspond to the flux density weighted velocity of the associated feature. The size of the data points (relative to each source) are scaled to the integrated flux density of the strongest maser channel of the associated feature.}
  \label{fig:G305_spots}
\end{figure}

In Figure~\ref{fig:G305_spots} we present the first VLBI maps of the~6.7--GHz methanol maser emission towards G\,305.200$+$0.019, G\,305.202$+$0.208 and G\,305.208$+$0.206. These have been produced from the~2013~November epoch of observations. High--resolution~(1.8$\times $1.4~arcsec$^2$ beam size) images of the~6.7--GHz methanol masers in G\,305.202$+$0.208 have previously been made by \citet{Phillips+98}, and \citet{Norris+93} and \citeauthor{Phillips+98} have presented maps of the~6.7--GHz methanol masers in G\,305.208$+$0.206. In producing the spot maps we grouped the emission for each source which was present across contiguous spectral channels into a single maser feature. Each feature typically represents emission from between~4~--~20~spectral channels~(0.22~--~1.10~\kms ), and the points in the figures are a flux density weighted average in position and velocity. The methanol emission for these sources is measured to span between~190~--~1000~AU, which is consistent with measurements from other VLBI observations of methanol maser emission \citep[e.g.][]{Brunthaler+09,Bart+08}. The errors in the relative positions between the features in each map are~$<<$1~mas and the image noise is between~10~--~20~mJy~beam$^{-1}$.

\subsubsection{G\,305.200$+$0.019}
\label{sec:g-305.200+0.019}

We find the 6.7--GHz methanol maser emission associated with G\,305.200$+$0.019 has eight distinct velocity components in Figure~\ref{fig:G305_spots}~(top panel). The emission spans~0.50\arcsec \/ in the north--south direction and~0.12\arcsec \/ along the east--west direction. Except for the feature at~$-$31.96~\kms , there is a velocity gradient in the emission with the seven features which form a simple and linear structure towards the north of the plot. This structure spans~0.06\arcsec \/ (averaged along either axis), which corresponds to to~245~AU at a distance of~4.1~kpc~(from Table~\ref{tab:parResII}). The lone feature at~$-$37.07~\kms \/ is offset from the centre of the image field by~0.51\arcsec . The maser spot which was used for parallax determination is associated with the feature with an integrated flux density of~30.0~Jy at~$-$33.15~\kms . In Figure~\ref{fig:G305_200_spec}, we identify a feature at~$-$38.5~\kms \/ for G305.200$+$0.019 which was undetected in \citet{Green+12a}. We were not able to identify this feature from our~1.2~arcsec$^2$ image cube, and conclude that it is either resolved on VLBI baselines, or may be associated with G\,305.199$+$0.005.

\citet{Hindson+12} derive the properties of the embedded high--mass star for the candidate UC\,\ionhy \/ region associated with G\,305.200$+$0.019 and classify it as a~B\,1 source. Using their results along with the equations from \citet{Panagia+78}, we estimate the electron density~$n_e$, mass of ionized hydrogen~$M_{\mbox{\scriptsize \ionhy }}$, the Lyman continuum photon flux~$N_L$ in Table~\ref{tab:starTypeG305}. From this we find the~(lower limit) mass of the ionizing source responsible for the observed Lyman flux to be a~B\,0.5~type star.

Using the distance of~4.1~kpc for the \emph{IRAS} source identified by \citet{Walsh+97}, we find that G\,305.202$+$0.019 has a luminosity of~$L = 28.3\times 10^4$~\lsol . This corresponds to a source with an~O\,6~--~5.5 spectral type \citep{Panagia+73}, which is in contrast to the classification in Table~\ref{tab:starTypeG305}. The disparity can be explained by high--mass stars forming in clusters, and the derived spectral type from the arcminute resolution \emph{IRAS} observations is likely to indicate the total luminosity of the cluster as if it were produced by a single high--mass source.

\begin{table}
\caption{Physical parameters of G\,305.200$+$0.019 from \citet{Hindson+12} adjusted to a preferred distance of~4.1~kpc.}
\centering
\begin{tabular}{cccccc}
\hline
  $n_e$            & $M_{\mbox{\mbox{\tiny \ionhy}}} $ & $\log ~N_L$ & $M_*$     & Spectral \\
 (cm$^{-3}$)       & (\msol )                       & (s$^{-1}$)  & (\msol )  & type   \\
\hline
\hline
$0.56 \times 10^4$ & $38.7 \times 10^{-4}$          & 46.19       & 14.9     & B\,0.5  \\
\hline
 \end{tabular}
  \label{tab:starTypeG305}
\end{table}

\subsubsection{G\,305.202$+$0.208}
\label{sec:g-305.202+0.208-g}
The middle panel of Figure~\ref{fig:G305_spots} shows the emission associated with G\,305.202$+$0.208 consists of four features with a simple structure spanning~0.08\arcsec \/ in the~east--west~direction, corresponding to~330~AU at a distance of~4.1~kpc (from Table~\ref{tab:parResII}). While the structure seen here is similar to that in \citet{Phillips+98}, it appears that some weak emission to the~north--east and~north--west has fallen below the detection limit of our observations or has been resolved by VLBI. However, the feature at~$-$43.26~\kms \/ was not detected by \citet{Phillips+98}. The maser spot which was used for parallax determination is associated with the feature with an integrated flux density of~22.5~Jy at~$-$44.03~\kms .

\citet{Walsh+97} derive a luminosity of~$20.9\times 10^4$~\lsol \/ for the \emph{IRAS} source associated with G\,305.202$+$0.208. (\citet{Walsh+99} show that the~IR~emission is associated with G\,305.202$+$0.208 and not G\,305.208$+$0.206 as presented in \citet{Walsh+97}.) Using the updated distance of~4.1~kpc to G\,305.202$+$0.208, we find that the luminosity is~$16.6\times 10^4$~\lsol , which would correspond to an~O\,6.5~spectral source, if solitary, but is likely to be indicative of the cluster luminosity.

\subsubsection{G\,305.208$+$0.206}
\label{sec:g-305.208+0.206}
G\,305.208$+$0.206 shows the most complex distribution of emission of the sources in Figure~\ref{fig:G305_spots}~(bottom panel), consisting of ten features. The main cluster forms a ring like structure with diameter of~$\sim $0.22\arcsec , corresponding to~900~AU at at distance of~4.1~kpc (from Table~\ref{tab:parResII}). There is also a feature at $-$35.55~\kms \/ offset to the~north--west by~0.38\arcsec \/ from the centre of the image field. The emission does not appear to follow a simple or clear velocity gradient, however there is a general decrease in velocity toward the south. \citet{Phillips+98} identify the strongest feature of this source as a single spot, however we have resolved the emission into two features at~$-$38.51 and~$-$38.09~\kms . There is also weak emission at~$-$46~\kms \/ in \citet{Phillips+98}, in the region between the~$-$36.15~and~$-$34.04~\kms \/ features which is below the detection limit of our observations. The integrated flux density of the peak feature at~38.51~\kms \/ is~122.3~Jy.

\section{Conclusion}
We have obtained parallaxes for methanol masers associated with the G\,305.2 region. The parallax of G\,305.200$+$0.019 is measured to be~0.21$\pm $0.06~mas and the parallax of G\,305.202$+$0.208 is measured to be~0.42$\pm $0.13~mas. We combine these to obtain a variance weighted average parallax of~0.25$\pm $0.05~mas, corresponding to a distance of~4.1$^{+1.2}_{-0.7}$~kpc to the~G\,305.2 region. We find the~(near) kinematic distance, using the latest Galactic parameters, to the~$^{13}$CO~molecular clump associated with these masers to be~4.3$^{+2.2}_{-1.4}$~kpc. While not inconsistent with the parallax distance, the kinematic distance has a larger uncertainty and provides evidence of the unreliability of this technique on providing distances for the study of~HMSF.

VLBI observations of maser emission associated with HMSFRs are providing insights into the spiral structure of the Milky Way galaxy. These measurements are predominantly from northern hemisphere observations and do not probe the Galactic structure in the forth quadrant. The LBA is the only VLBI instrument in the southern hemisphere which is currently providing parallax distances to methanol maser sources in this region. Our LBA parallax results from G\,305.200$+$0.019 and G\,305.202$+$0.208 allow us to place the G\,305.2 region in the Carina--Sagittarius arm, thereby extending the Galactic azimuth range of the sources in this arm by~$\sim $40$^\circ $. From this, we have revised the pitch angle of the Carina--Sagittarius arm to~$\psi = 19.0 \pm 2.6^\circ $. The broader azimuth range allows the pitch angle to be determined with greater confidence, demonstrating the role of southern hemisphere observations in determining the spiral structure of the Milky Way.\\

The LBA is part of the Australia Telescope National Facility which is funded by the Australian Government for operation as a National Facility managed by CSIRO and the University of Tasmania. We thank the referee~(Dr~Anita Richards) for helpful comments in reviewing this paper.

\bibliography{G305_ref}
\end{document}